\pdfoutput=1 %
\documentclass[10pt]{article}
\usepackage[a4paper,left=2.54cm,top=2.54cm,right=2.54cm,bottom=2.54cm]{geometry}
\usepackage{fancyhdr}
\usepackage{afterpage}
\usepackage{setspace}

\newfont{\toto}{msbm10 at 12 pt}
\newfont{\ithd}{cmr9}

\newcommand{\figu}[1]{\ref{fig:#1}}
\newcommand{\lafi}[1]{\label{fig:#1}}

\usepackage{amsmath,amsthm,amsfonts,amssymb}
\usepackage[pdftex]{graphicx} 
\usepackage[T1]{fontenc}
\usepackage[inline]{enumitem}
\usepackage[overload]{empheq} 
\usepackage{physunits} 
\usepackage{nicematrix}
\usepackage{booktabs}
\usepackage[
  colorlinks,
]{hyperref}
\usepackage{caption}
\usepackage{subcaption}

\renewcommand\vec[1]{\boldsymbol{#1}} %
\newcommand\myfullwidth{159.2mm}

\usepackage{titling}
\pretitle{\vspace{-4em}\begin{center}\LARGE}
\posttitle{\end{center}\vspace{-1em}}
\preauthor{\begin{center}\large}
\postauthor{\end{center}\vspace{-3em}}

\title{
\bf Coupling Machine Learning Local Predictions with a Computational Fluid Dynamics Solver to Accelerate Transient Buoyant Plume Simulations
}
\author{
      C. Caron$^{*,**}$, P. Lauret$^{*}$ and A. Bastide$^{*}$\\
      Corresponding author: clement.caron@univ-reunion.fr
      $^{*}$ {Department of Sustainable Built Environment, PIMENT lab, University of Reunion}, {La Réunion}, {France}.\\
      $^{**}$ {Research \& Development team}, 
      {INTEGRALE Ingénierie}, {Saint-Gilles les Hauts}, {La Réunion},
      {France}.
}
\date{}

\begin{document}

\maketitle
\afterpage{\fancyhead{}}

\centerline{
\begin{minipage}[t]{150mm}
{\bf Abstract:} 

Data-driven methods demonstrate considerable potential for accelerating the inherently expensive computational fluid dynamics (CFD) solvers.
Nevertheless, pure machine-learning surrogate models face challenges in ensuring physical consistency and scaling up to address real-world problems. %
This study presents a versatile and scalable hybrid methodology, combining CFD and machine learning, to accelerate long-term incompressible fluid flow simulations without compromising accuracy.
A neural network was trained offline using simulated data of various two-dimensional transient buoyant plume flows.
The objective was to leverage local features to predict the temporal changes in the pressure field in comparable scenarios.
Due to cell-level predictions, the methodology was successfully applied to diverse geometries without additional training.
Pressure estimates were employed as initial values to accelerate the pressure-velocity coupling procedure.
The results demonstrated an average improvement of $94\%$ in the initial guess for solving the Poisson equation.
The first pressure corrector acceleration reached a mean factor of $3$, depending on the iterative solver employed. 
Our work reveals that machine learning estimates at the cell level can enhance the efficiency of CFD iterative linear solvers while maintaining accuracy.
Although the scalability of the methodology to more complex cases has yet to be demonstrated, this study underscores the prospective value of domain-specific hybrid solvers for CFD.

\vskip0.2cm
{\it Keywords:} Computational fluid dynamics, Machine learning, buoyant plume, incompressible flow, Poisson equation. \\
\end{minipage}
}
\vskip0.5cm

\section{Introduction}

Computational fluid dynamics (CFD) techniques enable the simulation of fluid flow, which is valuable for a wide range of scientific challenges.
This flexible numerical approach can model various physical phenomena across different time and space scales based on the governing equations.
As a result, CFD finds applications in many areas, such as aerodynamics, turbomachinery, hydrology, chemical processes, meteorology, and buildings \cite{ferziger_computational_2020, versteeg_introduction_2007}.
However, the computational time required to solve unsteady large-scale real-world problems remains a significant bottleneck.
Despite the development of efficient numerical approaches and the growth of computational power, CFD is inaccessible for numerous applications \cite{runchal_cfd_2020, morozova_feasibility_2020}.
In this context, the CFD community is paying greater attention to machine learning algorithms to accelerate solvers by creating cost-effective models 
\cite{runchal_cfd_2020, brunton_machine_2020, vinuesa_enhancing_2022, calzolari_deep_2022, zehtabiyan-rezaie_data-driven_2022-1}.
The machine learning field encompasses various algorithms that extract valuable information from data, leading to applications such as pattern recognition or surrogate modeling \cite{hastie_elements_2009}.
Data-driven models can provide fast predictions, rendering them a desirable alternative to high-fidelity physics-based simulations.
Nevertheless, further research is required to identify effective ways to combine CFD with machine learning to speed up simulations while maintaining accuracy \cite{10124114}.

Although machine learning algorithms are not novel, they have substantially advanced recently.
A combination of factors, including abundant data, more capable hardware, increased computational power, and the development of efficient algorithms, has revealed the potential of machine learning.
Deep learning, which relies on deep neural networks, has gained significant attention due to its demonstration of state-of-the-art capabilities in addressing numerous intricate challenges, such as image or speech recognition \cite{bengio_deep_2017, lecun_deep_2015}.
The \emph{scientific machine learning} domain aims to bridge the gap between machine learning and scientific computing.
Concerning CFD, a primary focus is the development of data-driven models that achieve a more favorable balance between accuracy and computational cost \cite{vinuesa_enhancing_2022, 10124114}.  
However, several challenges must be overcome, including model generalizability, interpretability, scalability, and data efficiency \cite{zehtabiyan-rezaie_data-driven_2022-1, 10124114, faroughi_physics-guided_2024}.
Therefore, the research effort is shifting toward frameworks that enforce the underlying physics \cite{faroughi_physics-guided_2024}. 
For example, physics-informed neural networks \cite{raissi_physics-informed_2019} and neural operators \cite{li_neural_2020} hold significant potential for accelerating traditional scientific computing.
Additionally, \emph{hybrid approaches}, which combine physics-based solvers with machine learning, are considered a more practical strategy than completely replacing traditional numerical methods \cite{10124114}.
While hybrid approaches may be more complex to develop and yield less impressive computational advantages than data-driven black-box models, they could guarantee essential physical consistency.

\medskip

The existing literature suggests several strategies for hybridizing numerical and neural architectures for CFD. 
For instance, it is advisable to employ machine learning models to refine a cost-effective physical approximation rather than predict an absolute value \cite{list_how_2024}.
In this context, super-resolution techniques have been employed to reconstruct high-fidelity fields from low-fidelity data obtained, e.g., from coarse-grid simulations \cite{kochkov_machine_2021-1}.
However, despite model accuracy, transient simulations present an additional challenge due to the inevitable accumulation of errors when making consecutive data-driven predictions \cite{jeon_finite_2022, jeon_residual-based_2024, pedro_souza_de_oliveira_coupling_2022, peng_fourier_2024}.
While unrolling over multiple time steps in training enhances model performance, a more robust hybridization is required for long-term predictions \cite{list_how_2024, kochkov_machine_2021-1, ajuria_illarramendi_performance_2022}.
For example, a hybrid methodology alternating traditional CFD and machine-learning-based time series has been demonstrated to control this error effectively \cite{jeon_residual-based_2024}.
 
An alternative approach to achieve consistent results is to use the model output as the initial condition for a numerical solver, ensuring convergence constraints \cite{obiols-sales_cfdnet_2020}. 
Providing better initial guesses for costly solver sub-components, such as the solution of the pressure Poisson equation for incompressible fluids, is also decisive.
Thus, data-driven predictions have demonstrated effectiveness as initial values for traditional iterative methods used in solving linear systems \cite{ajuria_illarramendi_performance_2022, chen_machine_2022, zhang_hybrid_2022}.
However, these studies involved complex deep learning architectures predicting a field of interest for the entire domain simultaneously, which may struggle to scale up to real-world cases.

Concurrently, some studies have concentrated on developing \emph{local approaches}. %
These strategies entail training models to map a local flow description to a local prediction.
Consequently, models are typically simpler as they operate within a constrained dimensional space.
Also, they can be applied to any domain size at inference time.
Local approaches are therefore promising for improving scalability, generalizability, and data efficiency 
\cite{kochkov_machine_2021-1, jeon_finite_2022, jeon_residual-based_2024, pedro_souza_de_oliveira_coupling_2022}.

In light of the aforementioned readings, the next step is to combine local and hybrid methodologies to cumulate their benefits for transient flow simulations \cite{sousa_application_2024, sousa_enhancing_2024}.
In particular, it would be crucial to examine whether local approaches can effectively accelerate the various iterative methods employed for solving the Poisson equation.

\medskip

To address this research gap, this study presents a hybrid methodology combining machine learning and CFD to accelerate traditional solvers for long-term incompressible fluid flow simulations.
The framework is designed to be versatile, scalable, accurate, robust, and data-efficient.
We used two-dimensional (2D) buoyant plume simulations to evaluate this strategy.
Specifically, we implemented a local approach to predict the pressure field.
A fully connected neural network was trained offline to map physics-based cell-level features to the evolution of the cell pressure between consecutive time steps.
Then, the predicted field was utilized as an initial guess to solve the Poisson equation.
We compared the performance of the hybrid approach using several state-of-the-art iterative solvers.
The evaluation was performed on various geometries and initial conditions without additional training.

This paper is structured as follows. Section \ref{sec:plume_num} presents the buoyant plume problem and the generic numerical method to approximate the solution.
Then, the hybrid solver methodology, combining machine learning and CFD, is described in section \ref{sec:ml_solver}.
Next, the numerical experiment designed to assess the feasibility of our strategy is detailed in section \ref{sec:expe}.
The test case, the numerical setup, and the model training procedure are explained.
Finally, the model performance and the acceleration provided by the hybrid method are discussed in section \ref{sec:results}.

\section{Buoyant plume numerical modeling}
\label{sec:plume_num}

This work focuses on the numerical simulation of a buoyant plume. 
Buoyancy effects occur due to density variations within a fluid under the influence of gravity, causing a lighter fluid to rise when surrounded by a denser fluid. 
Since temperature directly affects density, buoyancy drives numerous natural flows, such as those found in oceans and the atmosphere. 
A plume is generated when a fluid ascends, driven by buoyancy or momentum.
Plume flows are a subject of extensive study due to their complex behaviors, presenting challenges for simulation \cite{maragkos_large_2013, kondrashov_heater_2017}.

\subsection{Simple plume equations}

We start with the assumption of a single-phase, incompressible Newtonian fluid.
The non-isothermal Navier-Stokes equations, with the Boussinesq approximation, are used to model a simple plume.
The momentum \eqref{eq:bouss-mom}, continuity \eqref{eq:bouss-cont}, and energy \eqref{eq:bouss-energy} equations are written as follows in Cartesian coordinates:
\begin{align}[left=\empheqlbrace]
      & \frac{\partial{\vec{u}}}{\partial{t}} + (\vec{u} \cdot \nabla) \vec{u} = -\frac{1}{\rho_0}\nabla p + \nu \nabla^2 \vec{u} + \frac{\rho}{\rho_0}\vec{g} \label{eq:bouss-mom} \\
      & \nabla \cdot \vec{u} = 0 \label{eq:bouss-cont}                                                                                                                                                     \\
      & \frac{\partial{T}}{\partial{t}} + (\vec{u} \cdot \nabla) T =  \alpha \nabla^2 T \label{eq:bouss-energy}
\end{align}
where $\vec{u}$, $p$, and $T$ are respectively the fluid velocity, pressure, and temperature fields.
$\rho$ represents the variable fluid density, and $\rho_0$ is the constant reference density.
$\nu$ denotes the kinematic viscosity, $\alpha$ the thermal diffusivity, and $\vec{g}$ the gravitational field. %

The energy equation \eqref{eq:bouss-energy} represents the standard unsteady advection-diffusion heat transfer process without a heat source.
The incompressibility assumption holds under the condition that 
\begin{enumerate*}[label=(\roman*)]
      \item the velocities involved are low (i.e., subsonic flow) and that
      \item density variations remain moderate.
\end{enumerate*}
The latter point refers to the Boussinesq approximation, where density variations are considered only in the gravitational term of the momentum equation.
In addition, a linear dependence is maintained between density variation and temperature variation (Equation \ref{eq:bouss-rho}).
\begin{equation}
      \label{eq:bouss-rho}
      \rho_0 - \rho = \beta\rho_0(T-T_0)
\end{equation}
$\beta$ denotes the coefficient of thermal expansion, $T_0$ is the constant reference temperature related to $\rho_0$, and $T$ denotes the temperature related to $\rho$.
Thus, $\rho$ can be substituted in Equation \eqref{eq:bouss-mom}.

If the temperature differences are small enough, this modeling correctly captures the movements due to the buoyancy.
Thus, the system of governing equations \eqref{eq:bouss-mom}-\eqref{eq:bouss-cont}-\eqref{eq:bouss-energy}-\eqref{eq:bouss-rho}, 
along with initial and boundary conditions, provide us with a reasonable approximation for representing a simple plume flow.
The Boussinesq approximation can even extend to hot smoke control studies for small-scale fires \cite{rooney_strongly_1997}.
It should be noted, however, that this approximation may yield qualitatively wrong flows when temperature differences are significant \cite{ferziger_computational_2020}.

\subsection{Numerical procedure}
\label{subsec:num}

The buoyant plume problem is solved numerically using a traditional CFD procedure based on the finite volume method \cite{ferziger_computational_2020, versteeg_introduction_2007}.
The finite volume method is well suited for complex geometries and is inherently conservative, making it a popular choice in engineering.
First, the grid generation preprocessing step divides the computational domain into a finite number of contiguous cells (control volumes).
Then, the underlying equations are integrated over each cell. 
Next, the integral forms of the conservation equations are discretized into a system of algebraic equations.
Finally, a numerical procedure solves the algebraic equations (see, e.g., Figure \figu{piso}). %

The governing equations constitute an interdependent pressure-velocity system, solved using the Pressure Implicit with Splitting of Operators (PISO) algorithm \cite{ISSA198640, doi:10.1080/104077901753306601}.
The equations \eqref{eq:bouss-mom}-\eqref{eq:bouss-cont} are decoupled and solved sequentially for each time step.
Figure \figu{piso} presents a simplified flowchart of the PISO algorithm variant we utilized in this study. 
$t_n$ denotes the current time step, corresponding to the time index $n$.
In this paper, the quantities designated by a superscript $n-1$ represent those from the preceding time step.
For the sake of simplicity, the absence of a time index denotes the current time step. 
At time $t_n$, the principal PISO stages are as follows: %
\begin{enumerate}[label=(\Roman*)]

      \item \label{itm:predictor} \emph{Predictor step}\\
      The discretized momentum equation is solved using an estimate of the pressure field denoted $p^*$. 
      This calculation yields the intermediate velocity field $\vec{u^*}$, which does not satisfy the continuity equation (unless $p^*$ is correct).
      Usually, the previous time step field is employed as an initial guess for pressure: $p^*=p^{n-1}$.

      \item \label{itm:temp} The temperature field is computed by solving the energy equation.

      \item \label{itm:corr1} \emph{Corrector step 1}\\
      One can derive the pressure equation by applying the divergence operator to the rearranged discretized momentum equation and using the incompressibility hypothesis.
      This pressure-correction Equation is a Poisson equation that is particularly computationally expensive to solve.
      The pressure $p^*$ is corrected, followed by the velocity field $\vec{u^*}$. The latter is updated explicitly, ensuring the velocity adheres to the continuity equation.
      However, the momentum balance may not be satisfied after this step.
      Thus, PISO performs a second corrector step.

      \item \label{itm:corr2} \emph{Corrector step 2}\\
      The Poisson equation is assembled following the same procedure with the updated fields.
      Therefore, pressure and velocity are corrected again.
      After this second corrector, pressure and velocity are considered to be correct.

      \item The additional discretized transport equations, such as species, temperature, or turbulence quantities, are solved.
\end{enumerate}

\begin{figure}[h]
      \centering 
      \includegraphics[width=\myfullwidth]{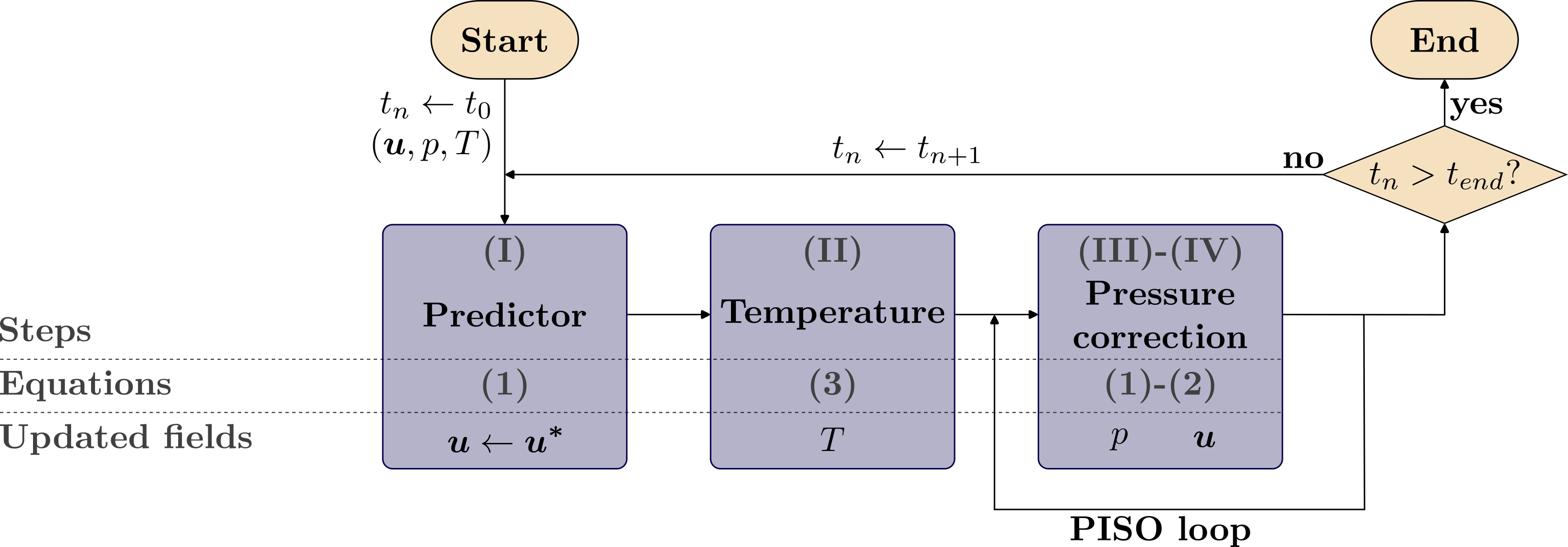}
      \caption{
            Simplified flowchart of the Pressure Implicit with Splitting of Operators (PISO) algorithm variant employed to address the buoyant plume case.
            We considered an incompressible flow with a fixed orthogonal mesh and without turbulence modeling.
      }
      \lafi{piso}
\end{figure}

\subsection{Iterative solvers}
\label{subsec:iter}

Each stage of the numerical procedure described above entails solving large-scale sparse linear systems whose size depends on the number of cells in the domain.
Solving the Poisson equation for pressure correction can consume up to 80\% of the computational time required to simulate incompressible fluid flow \cite{ajuria_illarramendi_towards_2020}.
Consequently, it is a significant challenge for CFD to accelerate linear solvers.
Standard direct methods, including Gauss elimination and LU decomposition, are unsuitable for solving large-scale CFD problems due to their excessive computational cost.
Therefore, CFD traditionally relies on iterative methods, which are generally much more efficient for solving such systems \cite{ferziger_computational_2020, versteeg_introduction_2007}.

Let us denote the discretized Poisson equation in matrix notation: $\mathbf{A}\mathbf{p}=\mathbf{b}$.
$\mathbf{A}$ is the sparse coefficient matrix, 
$\mathbf{b}$ is the right-hand side vector, and $\mathbf{p}$ is the unknown vector containing the pressure values for each cell.
Iterative methods start with an initial guess $\mathbf{p}^{(0)}$, which is improved iteratively through the repeated application of a cost-effective procedure.
$\mathbf{A}$ is decomposed, and the system is rearranged to define an iterative procedure
$\mathbf{M}\mathbf{p}^{(k+1)}=\mathbf{N}\mathbf{p}^{(k)}+\mathbf{c}$,
with $k$ as the iteration index, $\mathbf{M}^{-1}\mathbf{N}$ as the iteration matrix, and $\mathbf{c}$ as a constant vector.
Each iterative method employs a distinct decomposition strategy to achieve low-cost iterations and fast convergence. 
These linear solvers possess the advantage of memory efficiency and can be stopped when the desired tolerance is reached, which is convenient for adjusting the speed-accuracy tradeoff.
As the error $\mathbf{e}^{(k)} = \mathbf{p} - \mathbf{p}^{(k)}$ is unknown during the solving process, the residual $\mathbf{r}^{(k)}=\mathbf{b}-\mathbf{A}\mathbf{p}^{(k)}=\mathbf{A}\mathbf{e}^{(k)}$ is used to define the stopping criterion.
The number of iterations required to converge toward an acceptable level of accuracy cannot be predicted in advance.
However, it is closely related to the spectral radius of the iteration matrix and the initial guess.

Point-iterative techniques, such as Jacobi, Gauss-Seidel, or successive over-relaxation (SOR), 
are easy to implement but tend to exhibit a slow convergence rate for large systems. %
In particular, these methods encounter difficulties in eliminating low-frequency errors, which results in poor convergence performance on fine meshes.
Multigrid acceleration techniques have been developed to address this convergence issue.
Multigrid approaches use several resolution levels within the iterative process, following a cycle of coarsening and refinement. 
Point-iterative methods, called smoothers in this context, are efficient in eliminating high-frequency errors on a coarse grid, which correspond to the low-frequency errors of the finer grid.
Thus, multigrid methods exhibit accelerated convergence and represent state-of-the-art techniques for solving linear systems in CFD.
Krylov subspace methods, such as the Generalized Minimum Residual (GMRES) or the conjugate gradient method, are also popular in CFD.
Concerning conjugate gradient, the quadratic form of the linear system is used to convert it into a minimization problem.
Consequently, gradient-based minimization methods, such as the conjugate gradient method, can be executed. 
In practice, the problem is modified with preconditioners to improve convergence properties.
In the case of non-symmetric systems, the biconjugate gradients variant adapts the methodology.

This study aims to accelerate iterative solvers for the Poisson equation by providing a better initial guess $\mathbf{p}^{(0)}$. 
Although we do not provide any mathematical guarantee that an initial guess closer to the solution will significantly speed up the convergence, its impact can be critical.

\section{Hybrid solver methodology}
\label{sec:ml_solver}

This section introduces the hybrid CFD/machine-learning methodology designed to accelerate transient simulations without compromising accuracy.
The overall approach involves training a data-driven model to predict the future pressure field and then incorporating the prediction into the pressure-velocity coupling algorithm to expedite the pressure correction step.

\subsection{Prediction workflow}
\label{subsec:workflow}

The initial objective is establishing a workflow that estimates the pressure field $p^n:=p(\mathbf{x}, t_n)$ for each simulation time step $t_n$, with $\mathbf{x}$ as the position vector.
We aim to link the current state of the simulation, denoted as $S^n$, to the pressure field $p^{n}$.
$S^n$ encompasses all the available information at $t_n$ before the pressure correction step, such as the simulated quantities $\vec{u}(\mathbf{x}, t_{n-1})$, $p(\mathbf{x}, t_{n-1})$, or $T(\mathbf{x}, t_{n-1})$.
Note that $S^n$ may also include information from the previous time steps $\{t_0, \dots, t_{n-1}\}$, given that we have access to the complete time history of a simulation.
Figure \figu{workflow} depicts the prediction workflow for generating the pressure field.
The prediction strategy is founded upon three fundamental pillars. 
\begin{figure}[h]
      \centering
      \includegraphics[width=\myfullwidth]{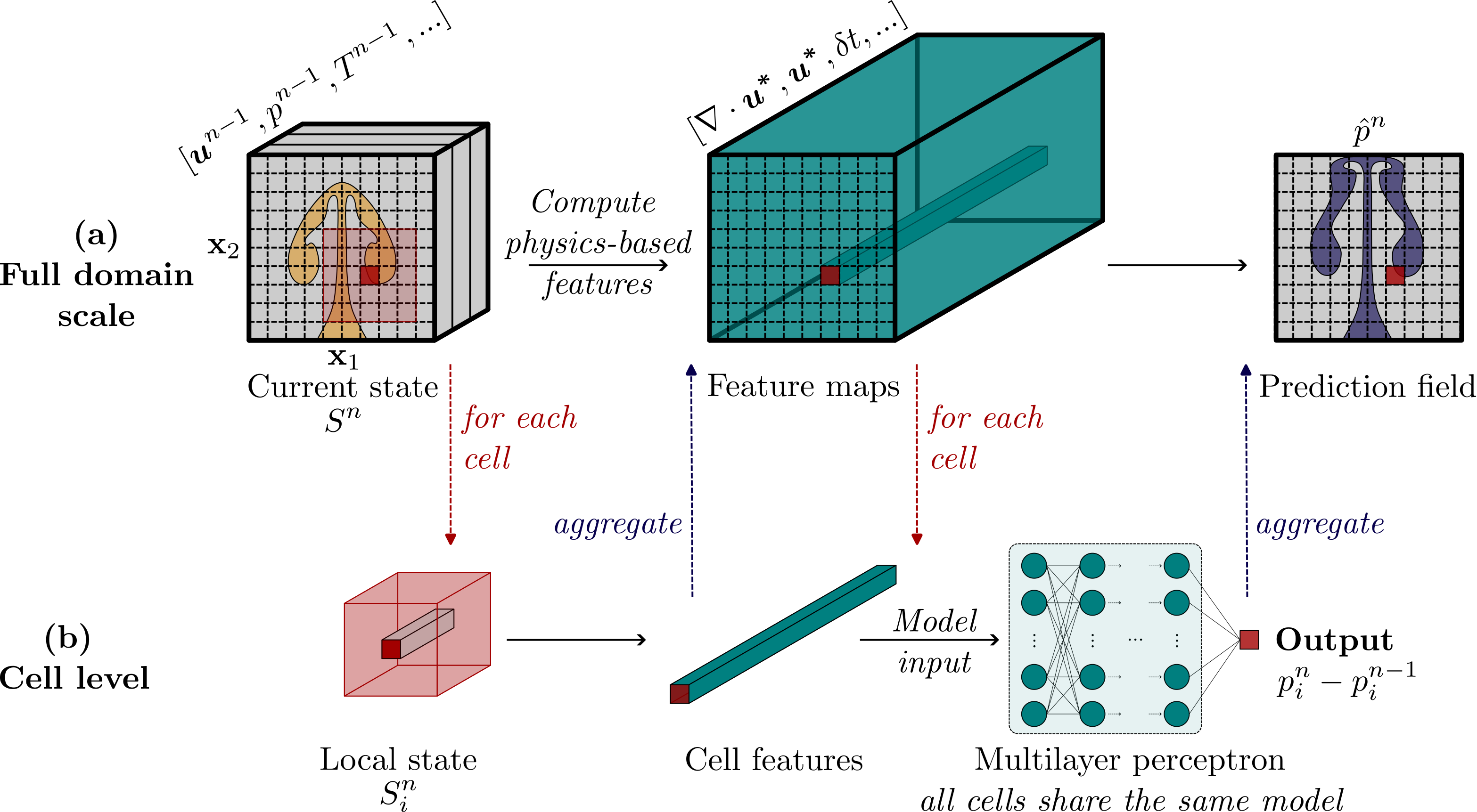}
      \caption{
            Illustration of the prediction workflow. 
            At a given time step $t_n$, we aim to predict the pressure field $p^n$.
            \textbf{(a)} Local physics-based features are computed for the entire domain, enabling the model to estimate the pressure field.
            \textbf{(b)} Features are calculated similarly for each cell $i$ of the domain. Some features incorporate information from the cell's neighborhood.
            The model infers the cell pressure variation from one time step to the next.
            }
            \lafi{workflow}
\end{figure}
\begin{enumerate}[label=(\roman*)]
      \item \emph{Cell-level predictions}. 
      The approach involves making predictions at the cell level rather than attempting to produce an estimate for the entire domain at once. 
      For all cells $i \in \{1,\dots, N_{cell}\}$ of the domain, the unique model maps a local state $S_i$ to a local pressure $p_i$.
      The local strategy is designed to scale to large domains without increasing model complexity.
      Additionally, the same model can make predictions for variable geometries.
      We also expect to enhance model generalizability by learning a local operator \cite{kochkov_machine_2021-1} and to train models with a small amount of data \cite{jeon_residual-based_2024}, i.e., data-efficient learning.
      While some studies coupling machine learning and CFD preferred a multi-scale approach for accurate predictions \cite{ajuria_illarramendi_performance_2022, chen_machine_2022}, 
      others have shown that a local approach can achieve reasonable accuracy \cite{jeon_residual-based_2024, pedro_souza_de_oliveira_coupling_2022}.
      Thus, we hypothesize that cell-level predictions can be sufficiently accurate for our hybrid solver.

      \item \emph{Pressure correction learning}.
      Instead of predicting the absolute pressure, the model learns to correct the pressure from the previous time step. 
      In other words, the model output is the local pressure temporal variation $p^n_i-p^{n-1}_i$.
      It has been proven that learning a correction can significantly improve model performance \cite{list_how_2024, jeon_finite_2022}.   

      \item \emph{Physics-based features}.
      The model inputs are handcrafted predictors based on physical quantities.
      For example, local features such as the divergence or the gradient of a quantity can be computed from the local state $S_i$.
      It should be noted that $S_i$ provides information about cell $i$ and its neighborhood, the size of which can be flexible.
      A few carefully chosen features may be adequate for accurate predictions.
      This strategy could lead to simple models, offering improved interpretability and faster inference.
      We posit that engineered features are worth considering before transitioning to more complex architectures, such as those typically utilized in deep learning.
\end{enumerate}

\subsection{Model learning strategy}

The data-driven model is trained using a supervised learning approach, illustrated in Figure \figu{supervised_ml}.
\begin{figure}[h]
      \centering
      \includegraphics{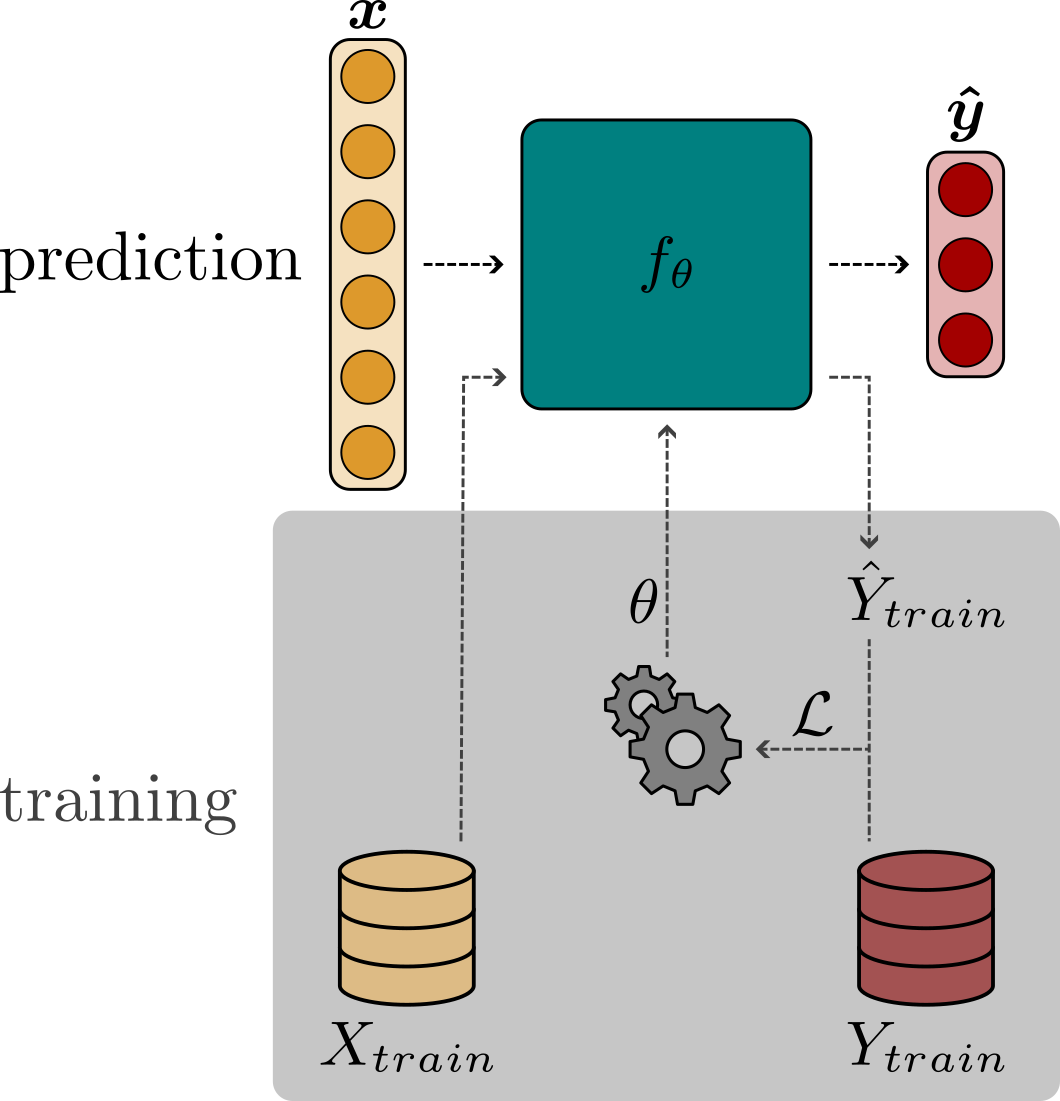}
      \caption{
            Supervised learning diagram. 
            The machine learning model $f_{\theta}$ provides a prediction $\vec{\hat{y}} \approx \vec{y}$ from an input vector $\vec{x}$.
            A database $(X_{train}, Y_{train})$ is used to train the model. 
            The training process aims to find optimal parameters $\theta$, according to some metric $\mathcal{L}$, to approximate the function $f$ mapping inputs $\vec{x}$ to outputs $\vec{y}$.
            }
            \lafi{supervised_ml}
\end{figure}
The training database contains CFD simulation examples.
For each simulation time step and domain cell, input/output couples $(\vec{x}, \vec{y})$ can be stored in matrices $(X_{train}, Y_{train})$.
$X_{train}$ is a $N_{train} \times N_{input}$ matrix, and $Y_{train}$ a $N_{train} \times N_{output}$ matrix.
$N_{train}$ denotes the number of points in the training set, $N_{input}$ is the input size, and $N_{output}$ is the output size.
Then, the model is trained to approximate the function $f: \vec{x} \mapsto \vec{y}$.
In this study, the input vector $\vec{x}$ corresponds to the physics-based features computed from the local state $S_i$, and the scalar output $y$ corresponds to the cell pressure correction $p^n_i-p^{n-1}_i$.
Machine learning algorithms, such as linear regressions, decision trees, or neural networks, are designed to learn the best possible approximation $f_{\theta}: \vec{x} \mapsto \vec{\hat{y}}$ of $f$ from data.
The parameters $\theta$ are automatically adjusted to minimize a loss function $\mathcal{L}$, which quantifies the discrepancy between the model predictions $\vec{\hat{y}}$ and the ground truth $\vec{y}$.

The objective is to develop a model that can be trained offline and provide satisfactory interpolation capabilities. 
Despite the time-consuming phases of data generation and training, the model remains advantageous as it can be deployed in a multitude of novel scenarios.
We believe it is preferable to specialize the model in a particular domain to achieve the highest possible prediction performance.
Indeed, developing a local machine-learning model for general-purpose CFD acceleration is a highly ambitious undertaking.
In our case, we concentrated on buoyant plume simulations, but this versatile hybrid approach could also be effectively applied to other domains.

\subsection{Model coupling approach}

In addressing transient simulations, it is anticipated that one-time-step machine learning predictions may not be sufficient to expedite long-term simulations.
Although a data-driven model may demonstrate strong predictive capabilities, 
it is likely to accumulate errors over successive time steps, which could result in a significant divergence from the physical flow 
\cite{jeon_finite_2022, jeon_residual-based_2024, pedro_souza_de_oliveira_coupling_2022, peng_fourier_2024}.
To guarantee long-term consistency, it becomes imperative to integrate data-driven outcomes within a traditional CFD solver. 

Given that the pressure correction steps \ref{itm:corr1}-\ref{itm:corr2} represent the most time-intensive phase for incompressible flows, efforts are directed toward enhancing the efficiency of this component.
Hence, we focus on pressure field predictions to build a Poisson equation solver surrogate model. 
Instead of substituting the Poisson equation iterative solver, we follow the idea of using model predictions as an initial guess, which has demonstrated encouraging results \cite{ajuria_illarramendi_performance_2022, chen_machine_2022, zhang_hybrid_2022,sousa_application_2024,sousa_enhancing_2024}.

In the PISO procedure (Figure \figu{piso}), the coupling strategy involves running the prediction workflow (Figure \figu{workflow}) for each time step before the pressure correction \ref{itm:corr1}.
The approach implies additional calculations for each time step, which must be considered when comparing the overall computing time. 
Nevertheless, these extra computations are intended to be rapid, independent, and scalable to any domain size. 

According to the specified coupling strategy, it could be argued that it is preferable to use the intermediate pressure correction (after the first corrector) as the model output rather than the final pressure (after the second corrector).
Nonetheless, the estimated outcome remains valid as the first corrector converges closely to the final pressure.
Additionally, this opens the door for the supplementary integration of a machine learning-based velocity correction, which would enhance the initial pressure-velocity coupling before pressure correction.

Thus, this coupling strategy guarantees the same level of accuracy as traditional CFD methods 
and is specifically designed to reduce the number of iterations for pressure correction, thereby reducing the overall computational time required.

\section{Experiment setup}
\label{sec:expe}

A numerical experiment was conducted to assess the potential of the methodology presented in section \ref{sec:ml_solver}.
The implementation of this innovative hybrid solver introduces, to the best of our knowledge, novel elements to the existing literature, including the following points:
\begin{enumerate}[label=(\roman*)]
      \item custom local physics-based features for data-driven pressure field estimation are evaluated,
      \item cell-level machine learning predictions are used as initial guesses for iterative linear solvers,
      \item the methodology is applied to a non-isothermal case with variable domain geometries and initial conditions without model retraining,
      \item the hybrid strategy performance is compared for several state-of-the-art iterative solvers.
\end{enumerate}

\subsection{Reference case}
\label{subsec:ref_case}

A reference case was defined and then adapted with diverse geometries and initial conditions to generate a training database. 
\begin{figure}[h]
      \centering 
      \includegraphics[width=\myfullwidth]{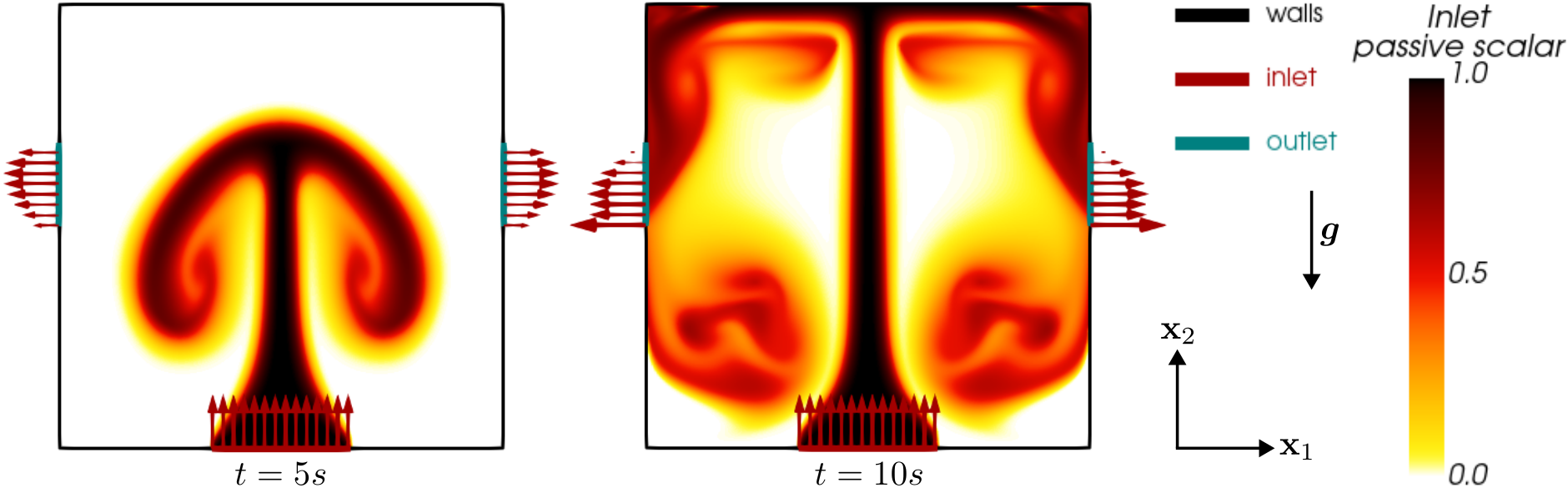}
      \caption{
            Reference case simulation at two different time steps.
            Inside the domain, colors represent the concentration of a passive scalar introduced at the inflow.
            The orthogonal projections of the velocity vectors at the boundaries are displayed to visualize the airflow rates.
            For this reference simulation, we set $L_{\mathbf{x}_1} = L_{\mathbf{x}_2} = 1 \meter$, $T_{inlet}=T_0+10$, and $U_{inlet}=0.125\meter.\Sec^{-1}$.
      }
      \lafi{ref_case}
\end{figure}
Inspired by similar studies \cite{ajuria_illarramendi_performance_2022, tompson_accelerating_2017}, the reference case was a 2D buoyant air plume in a $L_{\mathbf{x}_1} \times L_{\mathbf{x}_2}$ rectangular enclosure, as illustrated in Figure \figu{ref_case}. %
We simulated a one-square-meter box, closed with adiabatic walls and filled with air at a reference temperature of $T_0$. %
A unique inlet was positioned at the bottom of the box, through which warm air at a temperature $T_{inlet}$ was blown at a fixed uniform vertical velocity $U_{inlet}$. 
Two outlets were located on the vertical walls. 
The airflow could be directed either inwards or outwards at these openings according to the prevailing pressure conditions.
The reverse flow temperature was fixed to $T_0$.
The simulation was conducted for $10$ seconds, which allowed sufficient time for the plume to reach the ceiling due to the buoyancy force and for the emergence of various flow patterns (see Figure \figu{ref_case}).
To mimic air properties, we set $T_0=293.15\K$,
$\nu=15.06\times10^{-6}\meter^2/\Sec$, %
$\beta=3.43\times10^{-3}\K^{-1}$, the Prandtl number to $0.7$ and $g=9.81\meter/\Sec^2$.

\subsection{Physics solver setup}

We aimed to establish a robust physics solver configuration capable of solving 2D plume cases under various conditions with reasonable accuracy. 
This CFD solver was used to produce the training database, develop the hybrid solver, and serve as a benchmark for evaluating the hybrid approach. 
We were not striving to produce precise Direct Numerical Simulation (DNS) results, as it is unnecessary to demonstrate the feasibility of our methodology.

The open-source CFD software \texttt{OpenFOAM} (\texttt{v2312}) was employed to simulate the transient flows.
Specifically, we used the pressure-based solver \texttt{buoyantBoussinesqPimpleFoam}.
The solver was configured in PISO mode following the flowchart depicted in Figure \figu{piso}.
A uniform Cartesian grid comprising $256$ cells per meter was defined for spatial discretization.
Although the methodology employed did not impose such constraints on the mesh, this grid provided a simple starting point.
The boundary conditions employed are presented in Table \ref{t:bc}.
Simulations were run without turbulence modeling, which might introduce numerical errors. %
The use of stable numerical schemes was favored.
Therefore, a first-order backward Euler time integration was applied in conjunction with a first-order upwind scheme for the convective terms. %
The other spatial schemes relied on the default linear interpolation.
The temporal evolution of the flow was simulated using a variable time step.
Indeed, $\delta t$ was adjusted according to the maximum Courant number, set to $0.9$.
The linear systems were solved iteratively (see Section \ref{subsec:iter}) with a tolerance of $10^{-7}$ on the normalized residuals.
The multigrid solver \texttt{GAMG} with the Gauss-Seidel smoother was employed for the pressure, while conjugate gradient solvers were utilized for all other quantities.
\begin{table*}[h]
      \centering
      \NiceMatrixOptions{caption-above}
      \begin{NiceTabular}{lccc}[
            caption={Boundary conditions employed in \texttt{OpenFOAM} to solve the buoyant plume cases.},
            label={t:bc},
            tabularnote={
                  ${}^{\star}$ The total pressure was fixed to $\vec{g}\cdot\mathbf{h}$
            }
      ]
            \toprule
            & Velocity & Pressure & Temperature \\
            \midrule
            Wall & \texttt{noSlip} & \texttt{fixedFluxPressure} & \texttt{zeroGradient} \\
            Inlet & \Block[]{}{\texttt{fixedValue}} & \texttt{fixedFluxPressure} & \texttt{fixedValue}\\
            Outlet & \texttt{pressureNormalInletOutletVelocity} & \texttt{prghTotalPressure}${}^{\star}$
            & \texttt{inletOutlet}\\
            \bottomrule
      \end{NiceTabular}
\end{table*}

It is worth mentioning that, in \texttt{OpenFOAM}, the momentum equation is rearranged to include the hydrostatic pressure contribution $\rho(\vec{g}\cdot\mathbf{h})$ into the pressure gradient, which is numerically convenient.
$\mathbf{h}$ denotes the height vector in the opposite direction to gravity (our reference height is $0$).
Thus, by defining the alternative pressure 
$p_{rgh} =\frac{1}{\rho_0} (p - \rho(\vec{g}\cdot\mathbf{h}))$, Equation \refeq{eq:bouss-mom} becomes:

\begin{equation}
      \label{eq:momentum-of}
      \frac{\partial{\vec{u}}}{\partial{t}} + (\vec{u} \cdot \nabla) \vec{u} = -\nabla p_{rgh} + \nu \nabla^2 \vec{u} - (\vec{g}\cdot\mathbf{h})\nabla (\frac{\rho}{\rho_0})
\end{equation}
Indeed, based on the relationship
\begin{equation*}
      \begin{split}
            \nabla ( \frac{\rho}{\rho_0} \vec{g}\cdot\mathbf{h} ) &= \vec{g}\cdot\mathbf{h} \nabla (\frac{\rho}{\rho_0}) + \frac{\rho}{\rho_0} \underbrace{\nabla (\vec{g}\cdot\mathbf{h})}_{=\vec{g}} \\
      \end{split}
\end{equation*}
the gravity term  can be substituted in Equation \refeq{eq:bouss-mom}, leading to Equation \refeq{eq:momentum-of}.
Consequently, the target pressure in this experiment was not $p$, but $p_{rgh}$.
Nevertheless, for the sake of readability, we continue using the term pressure throughout this document. 

\subsection{Local features}
\label{subsec:features}

The machine learning model utilized physics-based local features as input to predict pressure evolution (see Figure \figu{workflow}).
These predictors were designed to be computationally inexpensive and to provide the model with insightful information for accurate predictions.
We deliberately opted for handcrafted features to keep the machine-learning process simple while providing baseline results.
However, identifying relevant features represents a significant challenge, contributing to the popularity of deep learning algorithms.
In light of this consideration, it is prudent to anticipate that representation learning algorithms have the potential to enhance our workflow by automatically extracting relevant features.

\begin{table*}[h]
      \centering
      \NiceMatrixOptions{caption-above}
      \begin{NiceTabularX}{\textwidth}{c|X[l]}[
            caption={List of cell-level features used as model input at time $t_n$.},
            label={t:features},
            tabularnote={
                  See Figure \figu{piso} and Section \ref{subsec:num} for PISO algorithm step numbering.
            },
            notes%
      ]
            \toprule
            Feature & Description\\
            \toprule
            $\vec{u^*}$ & \Block[t]{}{Intermediate velocity.\\Computed at the predictor step \ref{itm:predictor}.}  \\
            \midrule
            $T^*$ & \Block[t]{}{Intermediate temperature.\\Computed with the intermediate velocity $\vec{u^*}$ at step \ref{itm:temp}.} \\
            \midrule
            $p^*$ & \Block[t]{}{
                  Intermediate pressure.\\
                  The pressure estimate used for the predictor step \ref{itm:predictor} and as initial guess $\mathbf{p}^{(0)}$ for step \ref{itm:corr1}.
                  In this experiment $p^*= p_{rgh}^{n-1}$.}  \\
            \midrule
            $\delta t$ & \Block[t]{}{Time step between time $t_{n-1}$ and $t_n$.} \\
            \midrule
            $d_{inlet}$ & \Block[t]{}{Distance from cell to inlet.} \\
            \midrule
            $d_{outlet}$ & \Block[t]{}{Distance from cell to closest outlet.} \\
            \midrule
            $d_{wall}$ & \Block[t]{}{Distance from cell to closest wall.} \\
            \midrule
            $\delta \tilde{p}$ & \Block[t]{}{Filtered pressure.\\Difference between cell pressure and mean adjacent cell pressure.\\
                                                Here, $\delta \tilde{p} = p_{rgh}^{n-1} - \tilde{p}_{rgh}^{n-1}$ with $\tilde{p}$ the mean pressure on cell faces.
                                          } \\
            \midrule
            $\nabla \cdot \vec{u^*}$ & \Block[t]{}{Intermediate velocity divergence.} \\
            \midrule
            $\nabla \cdot (T^* \vec{u^*})$ & \Block[t]{}{$T^* \vec{u^*}$ product divergence field.} \\
            \midrule
            $\nabla \cdot (p^* \vec{u^*})$ & \Block[t]{}{$p^* \vec{u^*}$ product divergence field.} \\
            \midrule
            $\nabla T^*$ & \Block[t]{}{Intermediate temperature gradient.} \\
            \midrule
            $\nabla \cdot (\nabla T^*)$ & \Block[t]{}{Divergence of the temperature gradient.} \\

            \bottomrule
      \end{NiceTabularX}
\end{table*}

Table \ref{t:features} outlines the features selected for this experiment.
It is worth noting that all features were grid-independent, which preserved the flexibility of the methodology.
Most features depended only on the cell itself and the adjacent cells.
The distances to the boundaries, although requiring a more global perspective, were included because they might strongly impact the local fluid behavior. 
Considering the model target and the use of an adjustable time step, $\delta t$ emerged as an essential global feature.

\subsection{Dataset generation}
\label{subsec:data}

The training data were generated by running a series of $49$ simulations, all of which were variants of the reference case described in Section \ref{subsec:ref_case}.
The mesh size was maintained at $256$ cells per meter, and each simulation was $10\Sec$ long.

The following parameters were randomly assigned according to a uniform distribution.
\begin{itemize}
      \item Domain size $(L_{\mathbf{x}_1},L_{\mathbf{x}_2}) \in [1, 2]^2\meter$
      \item Inlet temperature $T_{inlet} \in [T_0+5, T_0+15]$
      \item Inlet velocity $U_{inlet} \in [0.1, 0.15]\meter.\Sec^{-1}$
      \item Inlet position and size.
      \item Outlet number (a maximum of one per boundary), positions, and sizes.
      \item Number of obstacles in the domain (up to three), positions, and sizes.
\end{itemize}
In addition, other rules were observed to ensure reasonable scenarios.
Obstacles, inlets, and outlets had minimum and maximum lengths, depending on the domain size. 
We also prevented the positioning of elements that were too close without overlapping.
Figure \figu{database} shows an extract of the generated database.

\begin{figure}[h]
      \centering 

      \includegraphics[width=\textwidth]{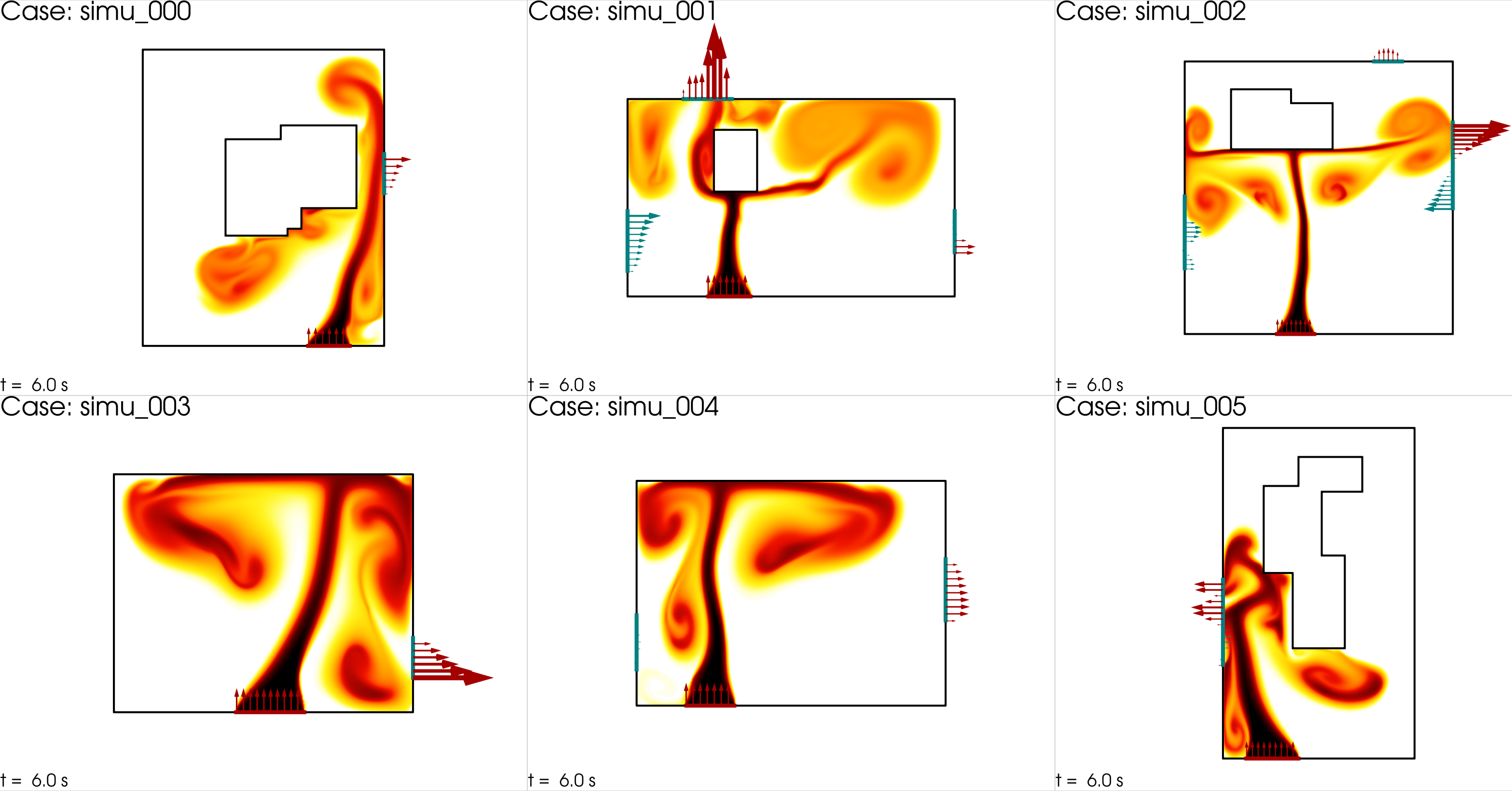}
      \caption{
            Database extract. Visualization of $6$ simulations at the same time step $t=6\Sec$.
            The domain size, initial conditions, and obstacles were randomly sampled.  
            The legend (colors, vectors) is the same as in Figure \figu{ref_case}.
      }
      \lafi{database}
\end{figure}

The data were only gathered after $1\Sec$ of simulation, which was arbitrarily chosen.
The rationale behind this was to avoid the initial phase of the simulation when the flow might exhibit some instabilities.
Consequently, using the model only after $1\Sec$ of simulation in the evaluation phase was also preferable.
It should be noted that this measure was introduced to facilitate model learning, although it might be unnecessary.

The cell-level procedure yields a large amount of data, as each domain cell and time step can produce a data point $(\vec{x}, \vec{y})$.
In our reference simulation, $65,536$ cells could generate over $90$ million data points within a $10\Sec$ simulation.
Thus, data points were sampled before aggregation in the database to manage a reasonable data volume while preserving a broad range of scenarios.
At a given time step, we set a $5\%$ probability of data generation. 
Data from $0.5\%$ randomly selected domain cells were collected during these time steps. 
Our database was segmented into three subsets, following established machine learning principles. 
The test set comprised $7$ complete simulations, while data from the remaining $42$ simulations were shuffled and split.
$80\%$ of the data was designated as the training set, with the other $20\%$ assigned to the validation set.  

\subsection{Model training}

The model was trained offline based on the generated training set described in Section \ref{subsec:data}.
Before training, standard preprocessing steps were carried out. 
First, the output pressure was modified by subtracting the pressure from the previous time step to learn a correction (see Section \ref{subsec:workflow}).
Then, the inputs and outputs were scaled, based on the training data, by removing the mean and scaling to unit variance (Equation \ref{eq:scaling}).
\begin{equation}
      \label{eq:scaling}
      \mathcal{T}(x) = \frac{x-\mu_{train}}{\sigma_{train}}
\end{equation}
$\mu_{train}$ and $\sigma_{train}$ represent, for the variable of interest $x$, the mean and the standard deviation of the training samples.
We denote the inverse transformation $\mathcal{T}^{-1}$ such that $\mathcal{T}^{-1}(\mathcal{T}(x))=x$.

Although the presented framework can be applied with any machine learning algorithm, this experiment focused on neural networks.
Indeed, neural networks have proven to be excellent universal approximators capable of handling large amounts of high-dimensional data.
Moreover, using such models in fluid mechanics has led to promising results, including accurate local predictions \cite{jeon_residual-based_2024, pedro_souza_de_oliveira_coupling_2022}.
Finally, neural networks provide interesting extensions for scientific machine learning applications (e.g., physics-informed neural networks, graph neural networks, or neural operators).

A neural network comprises artificial neurons that linearly combine input components and apply a final nonlinear activation function to produce a scalar output.
Subsequently, the neurons are organized into layers.
In this study, we chose the multi-layer perceptron (MLP) basic architecture as a starting point.
We anticipated that cell-level predictions do not necessitate a complex deep-learning architecture.
MLP layers are fully connected, as illustrated in Figure \figu{workflow}.
The optimization of the network involves adjusting the weights and biases of each unit.
These parameters, denoted by $\theta$, are optimized by minimizing a loss function with gradient-based optimization techniques.

The open-source machine-learning library \texttt{scikit-learn} \cite{scikit-learn} was employed to train the model, with the following implementation details. 
The MLP consisted of three hidden layers with $64$ units each ($64\times64\times64$).
It should be noted that the model architecture is typically fine-tuned as it can significantly impact performance.
However, the objective of this investigation was not to achieve optimal accuracy but rather to illustrate the viability of our methodology.
Therefore, this simple architecture was selected, which aligns with the optimal configuration identified in a comparable cell-level learning study \cite{jeon_finite_2022}.
The activation function employed was the rectified linear unit (ReLU).
The common squared error loss function described in Equation \eqref{eq:basic-loss} was minimized with the Adam optimizer \cite{kingma_adam_2017}.
\begin{equation}
      \label{eq:basic-loss}
      \mathcal{L}(\vec{y}, \vec{\hat{y}}) = || \vec{y} - \vec{\hat{y}} ||_2^2
\end{equation}
Other hyper-parameters were tuned with a grid search approach.
Thus, the tolerance for the optimization was set to $10^{-6}$, 
the learning rate to $5\times10^{-4}$, and the batch size to $1024$.
An $L_2$ regularization term with a coefficient of $10^{-4}$ was incorporated into the  Equation \ref{eq:basic-loss} loss function. 
The maximum number of iterations (i.e., epochs) was set to $150$.
Based on the validation set, an early-stopping strategy was implemented to prevent overfitting.

\subsection{Model incorporation in physics solver}
\label{subsec:model-in-solver}

We explored the feasibility of incorporating a machine learning model into a state-of-the-art CFD solver for research and industrial applications. 
Thus, the model was embedded into the \texttt{OpenFOAM} physics solver, which is implemented in C++.
In its current prototype stage, our hybrid solver performed model inference in Python, with data exchanges occurring through an inefficient file read/write process.
While a direct comparison of overall computation time would require a complete C++ implementation, we could still evaluate the model's impact on pressure correction.
To reduce the evaluation computational cost and assess the model's integration in varying scenarios, 
we used the model predictions at specific time intervals during the test simulations. 
Specifically, the model was activated every $10$ time steps, for a total of $1,045$ time steps across the $7$ test simulations.

The effectiveness of the hybrid solver was assessed using the same time steps with three iterative solvers, 
encompassing the main families outlined in Section \ref{subsec:iter}.
\begin{enumerate}[label=(\roman*)]
      \item \texttt{symGaussSeidel}. Symmetric Gauss-Seidel implementation, performing forward and reverse sweeps for each iteration.
      \item \texttt{PCG-DIC}. Preconditioned conjugate gradient, with simplified Diagonal-based Incomplete Cholesky preconditioner.
      \item \texttt{GAMG-GS}. Multigrid solver using a V-cycle, with Gauss-Seidel smoother.
\end{enumerate}
The same absolute tolerance was employed to solve the Poisson equation to ensure a fair comparison between the hybrid solver and the physics solver.
Convergence was reached when the residuals were below the $10^{-6}$ threshold, i.e., $||\mathbf{r}^{(k)}||_1=||\mathbf{b}-\mathbf{A}\mathbf{p}^{(k)}||_1 < 10^{-6}$.
It should be noted that the residual normalization procedure was deactivated to guarantee the same level of convergence, as this factor depends on the initial solution $\mathbf{p}^{(0)}$.

\section{Results and discussion}
\label{sec:results}

\subsection{Single-step predictions}
Since the loss function (see Equation \ref{eq:basic-loss}) does not provide any relative performance regarding model predictions, we define the skill score $\mathit{SS_{MSE}}$, which is expressed in Equation \ref{eq:ss}.
\begin{equation}
      \label{eq:ss}
      \mathit{SS_{MSE}} = 1 - \frac{\mathit{MSE}(Y_{val}, \hat{Y}_{val})}{\mathit{MSE}(Y_{val}, {Y}_{ref})}
\end{equation}
\begin{itemize}
      \item $\mathit{MSE}$ denotes the mean squared error between two vectors.
      \item $Y_{val}$ vector, with a size of $N_{val}$, represents the validation data ground truth before dataset preprocessing. It contains the target pressure $p_{rgh}$ for each data point.
      \item $Y_{ref}$ refers to the reference estimation for pressure, corresponding to the pressure at the previous time step $p_{rgh}^{n-1}$.
      \item  $\hat{Y}_{val}$ is the model prediction. The predicted pressure is given by $\hat{p}_{rgh} := p_{rgh}^{n-1} + \mathcal{T}^{-1}(\hat{y})$, for each validation point.
      The model raw prediction $\hat{y}$ is a scalar corresponding to the scaled alternative pressure variation for a given cell, i.e., $y=\mathcal{T}(p_{rgh} - p_{rgh}^{n-1})$.
\end{itemize}
This skill score is a quantitative measure of the performance of the model in comparison to a reference.
The reference is the pressure at the previous time step, the standard initial guess for pressure correction (see Section \ref{subsec:num}).
A positive skill score indicates that the model outperforms the reference for the selected metric ($\mathit{MSE}$).
$\mathit{SS_{MSE}}$ maximum value is $1$, or $100\%$.

\medskip 

\begin{figure}[h]
      \centering 
      \includegraphics[width=90mm]{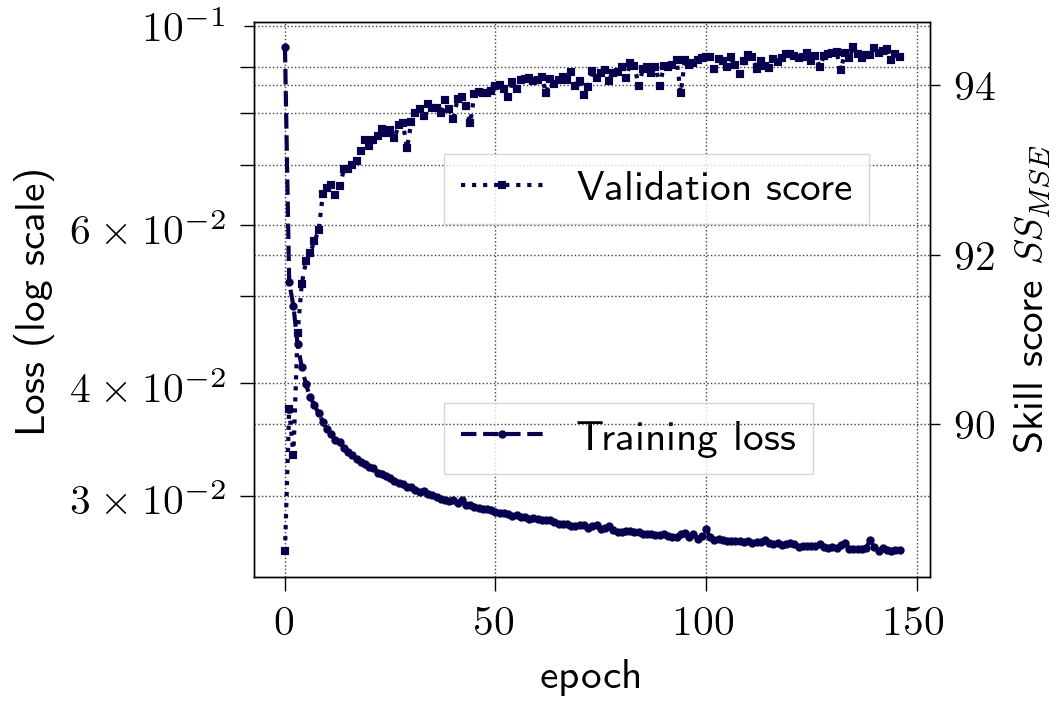}
      \caption{
            Model performance monitoring during the training phase.
            The decreasing curve (left Y-axis) shows the evolution of the loss function.
            The increasing curve (right Y-axis) indicates the skill score $\mathit{SS_{MSE}}$ on validation data.
      }
      \lafi{training}
\end{figure}

Figure \figu{training} exhibits the training phase monitoring.
The loss decreased from $9.48\times 10^{-2}$ at the end of the first epoch to $2.60\times 10^{-2}$ at the end of the training,
indicating that the learning procedure was effective.
In addition, the model demonstrated satisfactory performance on the validation set.
At the end of the training procedure, $\mathit{SS_{MSE}}$ reached $94.4\%$,
demonstrating that the mean squared error on the validation data was significantly lower with the machine learning model than with the reference estimation.
Although further improvements may be achievable, the considerable enhancements provided by these predictions over the reference should be noted.
The monotonic trend of the validation curve indicates that the selected model did not exhibit overfitting of the training data.
Nevertheless, the training was terminated before the maximum number of epochs was reached, as the skill score did not improve further.

To provide a more comprehensive evaluation of the model's performance, we computed the skill score $\mathit{SS_{Max}}$ with another discrepancy metric. 
The $\mathit{MSE}$ in Equation \ref{eq:ss} was replaced by the metric $\mathit{Max}(\vec{y}, \vec{\hat{y}})= || \vec{y} - \vec{\hat{y}} ||_\infty$.
The resulting value for $\mathit{SS_{Max}}$ was $75.6\%$.
The presented score indicates that the maximum error of the reference model was considerably larger than the maximum error of the proposed model on the validation set.
As the reference model is an acceptable physics-based approximator, 
the value of $\mathit{SS_{Max}}$ suggests that the machine learning model was stable in the sense that the predictions on unseen data did not deviate excessively from the ground truth.

\medskip 

At this juncture, we have demonstrated the viability of the prediction workflow and the learning strategy presented in Section \ref{sec:ml_solver}.
A single cell-level model could provide reasonable pressure field predictions for the next time step, which were more accurate than the reference when considering the 2-norm and the infinity norm.
The trained model exhibited notable performance on validation cases, indicating robust interpolation capabilities across diverse scenarios.

However, we still need to quantify the benefits that these predictions may bring to the iterative solvers (this point is discussed in Section \ref{subsec:hybrid-eval}).
Furthermore, it is important to acknowledge that the model was designed for single-time-step predictions and is likely inadequate for successive time steps without a hybrid strategy.
In addition, this initial demonstration is limited to the specified buoyant plume use case.

Further investigation is necessary to gain a more comprehensive understanding of the performance of the surrogate model.
First, it would be valuable to assess the model’s extrapolation capabilities.
For example, testing the model with larger domains or initial conditions beyond database distributions could yield valuable insights into its generalization performance.
In addition, it is crucial to determine whether the methodology can be extended to different grid types and resolutions.
Next, the learning approach could be applied to various quantities, such as the velocity, the temperature, or the pressure gradient.
Finally, optimizing the hyperparameters, including the model architecture, would enhance the prediction performance.

Nevertheless, this study provides encouraging evidence of the potential for local data-driven prediction capabilities.
This strategy could lead to simple machine learning architectures theoretically scalable to any domain size.
The adaptability of the methodology could enable the development of domain-specific models across a wide range of fields.

\subsection{Feature selection}

The physics-based features presented in Section \ref{subsec:features} were selected without a precise physical justification.
This raises the question of which quantities are relevant and whether other features might be more appropriate.
The first point was elucidated through the implementation of a feature selection algorithm. 

A forward sequential feature selection algorithm was employed to determine the relative importance of inputs.
The algorithm incorporates new features stepwise, selecting the one that yields the greatest improvement in the validation skill score $\mathit{SS_{MSE}}$ (Equation $\ref{eq:ss}$).
Therefore, the number of training instances to launch to rank $N_{input}$ features equals $\sum_{k=1}^{N_{input}}k$.
In our case, we identified $15$ scalar inputs, which led to the training of $120$ models.
Given the considerable time investment required for this strategy, we set a maximum of $50$ training iterations, which might allow for further improvements in some models. 
It should be noted that this greedy methodology does not provide the optimal combinations of features to maximize the skill score.
However, it does provide valuable insights with a reasonable algorithmic complexity.

\begin{figure}[h]
      \centering 
      \includegraphics[width=140mm]{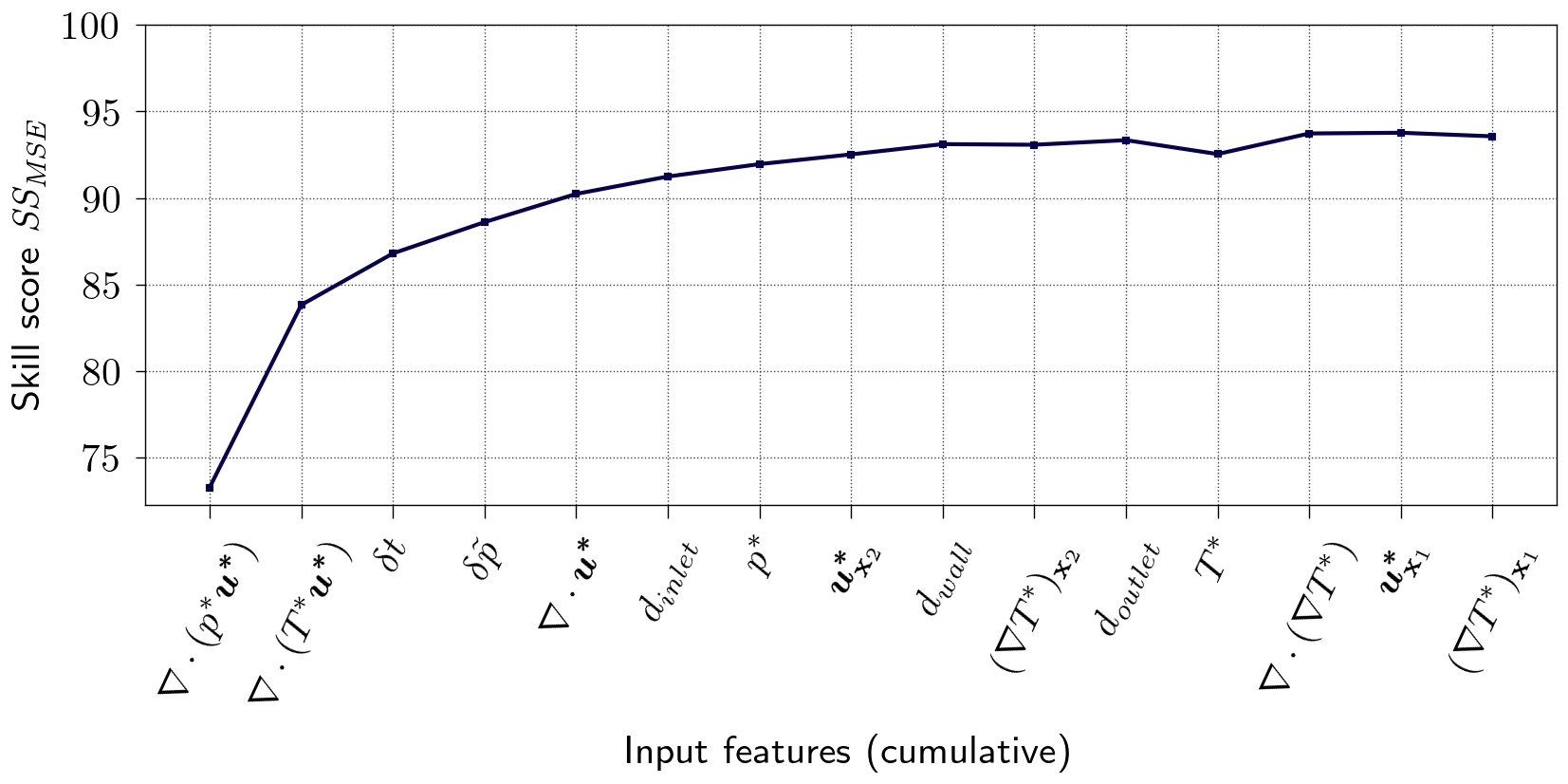}
      \caption{
            Forward sequential feature selection results.
            The skill score $\mathit{SS_{MSE}}$ was evaluated for each new feature added to the input vector.
      }
      \lafi{feature_select}
\end{figure}

Figure \figu{feature_select} depicts the evolution of the skill score  $\mathit{SS_{MSE}}$ as features were sequentially incorporated into the input vector, following the greedy procedure. 
First, we observe that a limited number of features can provide reasonable accuracy.
The skill score exceeded $90\%$ with just five input features.
Then, divergence fields were paramount in enhancing the model's accuracy.
Indeed, it seems plausible to suggest that the intermediate velocity divergence $\nabla \cdot \vec{u^*}$ is crucial, as it helps identify the continuity error when using the uncorrected pressure $p^*$ to solve the momentum equation.
Next, the relevance of the time step and distance features was established.
Finally, it seems that the raw quantities $T^*$, $p^*$, and $\vec{u^*}$ at the cell level provided little helpful information to improve the model.
It should be noted that, for vectors, the second component ($\mathbf{x}_2$) proved to be more pertinent than the first ($\mathbf{x}_1$), likely due to its alignment with the buoyancy force.
The findings may inform the development of parsimonious and fast models based on a few judiciously chosen predictors.  

Further feature engineering could lead to more relevant inputs, improving the model's performance.
In light of the temporal nature of the problem at hand, it would be appropriate to use time series as input so that the model may comprehend the dynamics of flow.
Additionally, it may be beneficial to increase the local state $S_i$ size, i.e., the model receptive field.
Indeed, a less restrictive view of the problem could lead to more informed and accurate predictions.
Finally, feature extraction techniques, including representation learning, constitute the logical next step in improving model features.

\subsection{Hybrid solver evaluation}
\label{subsec:hybrid-eval}

The acceleration factor $\eta(x) = \frac{x_{\text{CFD}}}{x_{\text{ML+CFD}}}$ is defined for each time step to assess the hybrid solver performance.
$x$ denotes the metric of interest:
\begin{itemize}
      \item $k_{\text{corr}1}$: the number of iterations to solve the first pressure corrector (step \ref{itm:corr1}, see Section \ref{subsec:num}).
      \item $\tau_{\text{corr}1}$: the processor time measured for the first pressure corrector.
      \item $\tau_{\text{corr}}$: the processor time measured for the entire pressure correction process (steps \ref{itm:corr1}-\ref{itm:corr2}, see Section \ref{subsec:num}).
\end{itemize}
Note that all the simulations were carried out under the same conditions with the same machine.
"CFD" refers to the physics solver using a traditional iterative method to solve the Poisson equation (see Section \ref{subsec:model-in-solver}):
Gauss-Seidel, preconditioned conjugate gradient, or multigrid.
The term "ML+CFD" refers to the hybrid solver presented in this paper, where the machine learning model prediction is used as an initial guess to solve the Poisson equation.

\begin{table*}[h]
      \centering
      \NiceMatrixOptions{caption-above}
      \begin{NiceTabularX}{\textwidth}{lX[c]X[c]X[c]}[
            caption={
                  Acceleration factor results. 
                  Comparison between the hybrid solver and the CFD solver. 
                  $k$ denotes the number of iterations, $\tau$ the processor time, $\text{corr1}$ the first pressure corrector, and $\text{corr}$ the entire pressure correction process.
                  Results are based on a sample of $1,045$ time steps distributed across $7$ test simulations. },
            label={t:factors}
      ]
            \toprule
            & \Block{1-3}{Iterative solver}\\
            \cmidrule{2-4}
            & \Block{}{ Gauss-Seidel\\ \texttt{symGaussSeidel}} &\Block{}{Conjugate gradient\\\texttt{PCG-DIC}} & \Block{}{Multigrid\\\texttt{GAMG-GS}} \\
            \cmidrule{2-4}
            $\eta(k_{\text{corr}1})$ mean value& $2.976$ &  $1.594$ & $1.264$\\
            $\eta(k_{\text{corr}1}) > 1$ ratio& $100.0\%$ & $90.4\%$ & $80.0\%$ \\
            \cmidrule{2-4}
            $\eta(\tau_{\text{corr}1})$ mean value & $2.967$& $1.580$ &$1.248$ \\
            $\eta(\tau_{\text{corr}})$  mean value& $2.461$& $1.349$ &$1.079$ \\

            \bottomrule
      \end{NiceTabularX}
\end{table*}

Table \ref{t:factors} summarizes the acceleration factors obtained with the hybrid approach.
All configurations were successfully accelerated on average, i.e., $\eta > 1$. 
As expected, the number of iterations required to solve the first Poisson equation was reduced.
The mean acceleration factors were approximately equal to $3.0$, $1.6$, and $1.3$ for Gauss-Seidel, preconditioned conjugate gradient, and multigrid solvers.
In the case of Gauss-Seidel, all the evaluation time steps were strictly improved, which is a highly encouraging outcome. %
This proportion fell to $90\%$ for the conjugate gradient and $80\%$ for the multigrid solver.
Therefore, the machine learning predictions offered a stable improvement to the iterative processes.
The boxplots in Figure \figu{it1} represent the number of iterations required for each iterative method. 
Significant distribution shifts are observed, favoring the hybrid method.

\begin{figure}[h]
      \centering
      \begin{subfigure}{0.32\textwidth}
          \centering
          \includegraphics[width=45mm]{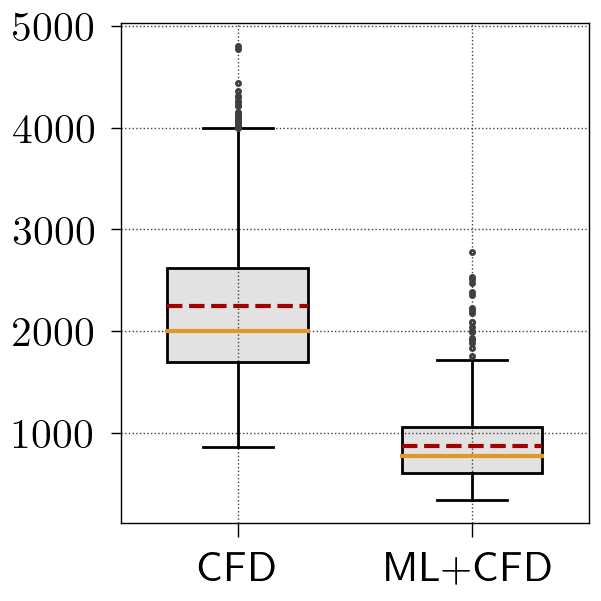}
          \caption{Gauss-Seidel}
          \lafi{it_gs}
      \end{subfigure}
      \hfill
      \begin{subfigure}{0.32\textwidth}
          \centering
          \includegraphics[width=45mm]{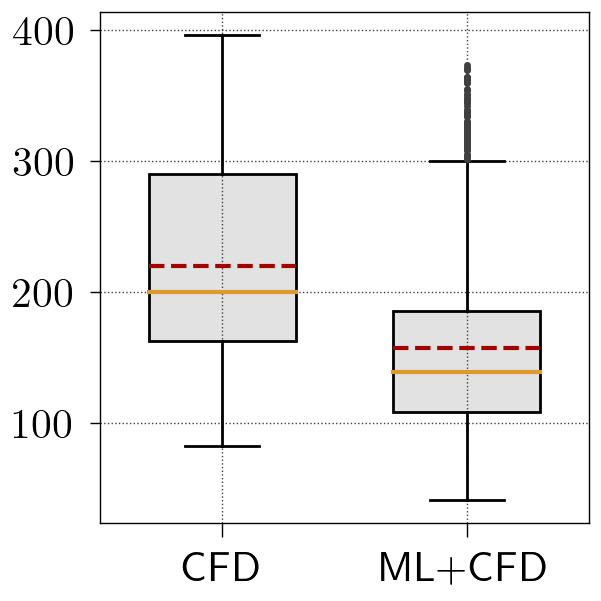}
          \caption{Conjugate gradient}
          \lafi{it_pcg}
      \end{subfigure}
      \hfill
      \begin{subfigure}{0.32\textwidth}
          \centering
          \includegraphics[width=45mm]{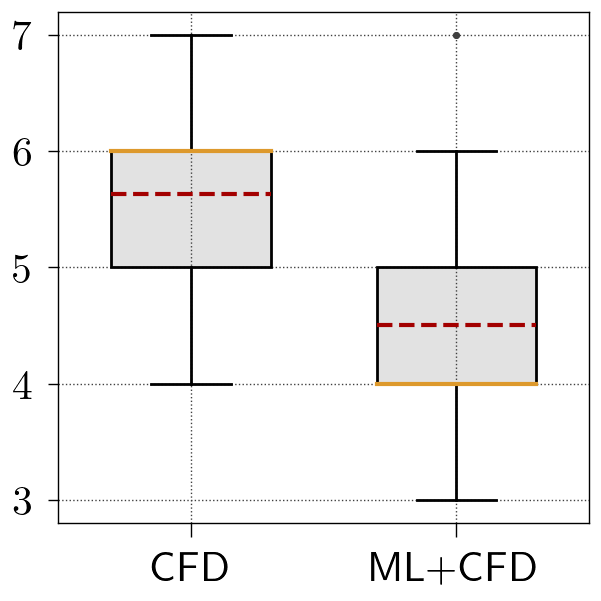}
          \caption{Multigrid}
          \lafi{it_gamg}
      \end{subfigure}
         \caption{
          Number of iterations to solve the first pressure correction.
          Comparison between the hybrid solver (ML+CFD) and the CFD solver. 
          Results are based on the same sample of $1,045$ time steps distributed across $7$ test simulations.
          Solid line: median. Dashed line: mean.
          }
         \lafi{it1}
  \end{figure}

Our hybrid methodology demonstrated a distinct advantage when using the Gauss-Seidel algorithm, a more moderate benefit for the preconditioned conjugate gradient techniques, and a less prominent advantage for the multigrid algorithm.
Although we cannot provide a straightforward explanation for this result, we notice that the approach performed better when coupled with iterative methods exhibiting a slow convergence rate.
As illustrated in Figure \figu{it1}, the number of required iterations to achieve convergence varied significantly depending on the iterative method employed.
Considering the baseline CFD solver, the mean number of iterations was approximately $2250$ for Gauss-Seidel, $220$ for the conjugate gradient, and $5.6$ for the multigrid approach.
Given the limited number of iterations required for the multigrid solver, larger-scale problems are worth considering. %
For example, an expanded domain, a three-dimensional (3D) problem, or a more refined grid might result in more favorable accelerations.

Regarding the computational time, a positive correlation was confirmed between the number of iterations $k$ and the computing time $\tau$.
As anticipated, the acceleration factors for the global pressure correction procedure $\eta(\tau_{\text{corr}})$ were less important than those for the first corrector $\eta(\tau_{\text{corr}1})$ since the second corrector was not enhanced.
This remark paves the way for further improvements, such as providing a more accurate initial estimate for the second corrector or enhancing the intermediate velocity $\vec{u^*}$ with machine learning to achieve a faster pressure-velocity coupling. %
\medskip 

This initial demonstration of the hybrid solver methodology is highly encouraging.
Our work demonstrates that cell-level machine-learning corrections of the pressure field can significantly accelerate CFD simulations while maintaining accuracy.
What may appear unexpected is that the model was not fine-tuned. A simple MLP architecture was employed, with basic physics-based local quantities as inputs.
In addition, simulations were run with a variable time step, which improves the approach flexibility.

A primary limitation is that the accelerations presented did not account for the time spent on model inference and feature computations, which must be performed for each cell in the domain. 
Consequently, the actual computational time savings are likely lower. 
Nevertheless, given the simple neural network architecture, the inference should be fast. 
Additionally, the model predictions and feature computations can be parallelized, and their associated cost is expected to increase linearly with the number of cells. 
Thus, the methodology should benefit problems above a certain size, provided that the acceleration factors are maintained and the iterative method employed does not scale linearly with the degrees of freedom \cite{sousa_enhancing_2024}.
However, this statement remains to be demonstrated for large-scale problems.

In this study, no comparison was made with respect to the tolerance set to $10^{-6}$ (see Section \ref{subsec:model-in-solver}).
The acceleration factors are anticipated to increase when the tolerance is reduced \cite{chen_machine_2022}, rendering the method even more attractive.
Nevertheless, it would be beneficial to corroborate this assertion within the context of our investigation.

Finally, only a single model was assessed in this experiment.
It is imperative to ascertain the influence of more accurate models on the observed acceleration.
The precise relationship between a reduction in quadratic error and the acceleration factors may not be immediately apparent.

Taking a step back, while we concentrated on the buoyant plume case, this coupling methodology is valuable for all incompressible simulations.
Local domain-specific models could emerge for a robust pressure-velocity coupling acceleration without any compromises in accuracy.

\section{Conclusion and future work}

This study introduced a hybrid solver methodology that combines a traditional CFD solver with a machine learning model.
This approach enables the acceleration of unsteady incompressible simulations while maintaining accuracy.
A generic workflow for data-driven cell-level predictions and a coupling strategy with the PISO algorithm were described to speed up the time-consuming pressure correction step.
The methodology entails the offline supervised training of a model, which can then be applied to various interpolated cases with disparate domain geometries and initial conditions.

The experimental results demonstrated the concept's applicability to non-isothermal 2D buoyant plume cases. 
In particular, we proved that cell-level pressure predictions can be employed as initial guesses to accelerate Poisson equation iterative solvers. 
The study is noteworthy for the simplicity of its neural network architecture, which relied on a limited number of handcrafted local physics-based features.
Finally, we conducted a comparative analysis of the impact of data-driven predictions on computing time across three distinct categories of state-of-the-art linear iterative solvers. 

In particular, a multi-layer perceptron enhanced the pressure initial guess by $94\%$ regarding the mean squared error. 
It was demonstrated that a $90\%$ enhancement could be achieved with only five predictors. 
Notably, the divergence fields associated with the intermediate quantities in the PISO algorithm were crucial features.
Finally, the hybrid strategy facilitated a reliable acceleration of the first pressure corrector in diverse test scenarios.
Observed acceleration factors ranged from $1.3$ with the multigrid solver to $3.0$ with the Gauss-Seidel algorithm.

Although the methodology is versatile and produced encouraging results, the scope of this study was limited to 2D buoyant plume cases with a uniform Cartesian mesh.
Furthermore, the time comparisons did not include the model inference or feature computations.

Future research should concentrate on a comprehensive time comparison involving larger cases, including 3D ones, to confirm the potential and scalability of the hybrid solver.
In addition, the extrapolation capabilities must be evaluated.
Assessing the methodology for unstructured meshes and turbulent flows would be beneficial in addressing increasingly complex applications.
Concurrently, the precision of the models can be improved to achieve greater acceleration.
Several avenues can be pursued, including model optimization (e.g., algorithm, architecture, hyperparameters) or feature engineering (e.g., temporal features, increased receptive field, representation learning).

This study constitutes a step toward integrating machine learning models into CFD solvers.
The hybrid methodology is versatile and could be applied to any incompressible flow.
We hope this study will inspire further investigations into coupling CFD solvers with local data-driven predictions,
leading to the development of reliable and scalable hybrid solutions for real-world challenges.

\bibliographystyle{unsrt} 
\bibliography{2024_ICCFD_paper}

\end{document}